\RequirePackage[dvips]{graphicx}
\documentclass[journal,onecolumn,  12pt, draftclsnofoot]{IEEEtran}
\usepackage{etex}
\usepackage{dcolumn}
\usepackage{epsfig}
\usepackage[cmex10]{amsmath}
\usepackage{amssymb}
\usepackage{amsthm}
\usepackage{amsfonts}
\usepackage{subcaption}
\usepackage{setspace}
\usepackage{bm}
\usepackage{cite}
\usepackage[svgnames]{xcolor} % cost before pstricks
\usepackage{pstricks,pst-node,pst-plot,pstricks-add}

\usepackage{algpseudocode}
\usepackage{algorithm}

\usepackage{tikz}

\usepackage{mathdots}
\usepackage{yhmath}
\usepackage{cancel}
\usepackage{color}
\usepackage{siunitx}
\usepackage{array}
\usepackage{multirow}
\usepackage{gensymb}
\usepackage{tabularx}
\usepackage{booktabs}
\usetikzlibrary{fadings}
\usetikzlibrary{patterns}
\usetikzlibrary{shadows.blur}
\usetikzlibrary{shapes}
\graphicspath{{figs/}}

\tikzset{every picture/.style={line width=0.75pt}} %set default line width to 0.75pt   
\input{Definition.def}

\newcommand{\figref}[1]{Fig.\,\ref{#1}}

\begin{document}

\date{}

\title{Optimal Communication-Computation Trade-Off in Heterogeneous Gradient Coding}
\author{Tayyebeh~Jahani-Nezhad, Mohammad~Ali~Maddah-Ali 
\thanks{
Tayyebeh Jahani-Nezhad is with  the  Department of Electrical Engineering, Sharif University of Technology, Tehran, Iran. (email:tayyebeh.jahaninezhad@ee.sharif.edu).
Mohammad Ali Maddah-Ali is with  the  Department of Electrical Engineering, Sharif University of Technology, Tehran, Iran. (email: maddah\_ali@sharif.edu).}
%\thanks{
%This paper has been presented in part at the  IEEE International Symposium on Information Theory (ISIT) 2018\cite{early}.
%}
 }
 \renewcommand\footnotemark{}
\maketitle

\begin{abstract} 
Gradient coding allows a master node to derive the aggregate of the partial gradients, calculated by some worker nodes over the local data sets,   with minimum communication cost, and in the presence of stragglers. In this paper, for gradient coding with linear encoding, we characterize the optimum communication cost for heterogeneous distributed systems with \emph{arbitrary} data placement,  with $s \in \mathbb{N}$ stragglers and $a \in \mathbb{N}$ adversarial nodes.   In particular, we show that the optimum communication cost, normalized by the size of the gradient vectors, is equal to $(r-s-2a)^{-1}$, where $r \in \mathbb{N}$ is the minimum number that a data partition is replicated. In other words, the communication cost is determined by the data partition with the minimum replication, irrespective of the structure of the placement. The proposed achievable scheme also allows us to target the computation of a polynomial function of the aggregated gradient matrix. It also allows us to borrow some ideas from approximation computing and propose an approximate gradient coding scheme for the cases when the repetition in data placement  is smaller than what is needed to meet the restriction imposed on communication cost or when the number of stragglers  appears to be more than the presumed value in the system design.
\end{abstract}

%\textbf{Index terms}$-$ Matrix Multiplication, Distributed computation, Approximation, Count-Sketch, Entangled polynomial codes, Coding for computing, Sparse Recovery.

\section{Introduction}
Distributed machine learning is known as an inevitable solution to the challenging task of training complicated models with large data sets \cite{dean2012large,4455,li2014communication}. In this scenario, the data set or the model parameters are distributed among several servers, that collaboratively perform the task of training or evaluating the model.  In one scenario, particularly popular for deep neural networks,  the model parameters are stored in a master node, and the data set is divided among worker nodes. Each worker node computes partial gradients of the model based on its local data set and returns back the results to the master node, where the results are aggregated to obtain the gradient over the whole data set. 

%Deep learning is one of the applications which use gradient computation as a key technique in the algorithm. Thus, as one scenario, assume that the model parameters are stored in a master node, and the data set is divided among worker nodes. Each worker node computes partial gradients based on its local data set and returns back the result to the master node, where the results are aggregated to obtain the gradient over the whole data set. 

Distributed machine learning runs into a list of challenges related to the rate of convergence, the cost of computation per node, privacy of the data,  existence of faulty nodes, etc. One of the major challenges is the cost of communication, or transferring the large amount of data, e.g., the gradient vectors~\cite{li2014communication}.  Another major challenge is to deal with stragglers, or slow servers, which dominates the speed of the computation~\cite{tail}.  On the other hand, some of the servers may be controlled  by the adversary, aiming to make the final results incorrect~\cite{blanchard2017machine, chen2017distributed,chen2018draco}. 
More recently, it is shown that coding can be effective in coping with stragglers and adversarial worker nodes as well as reducing communication cost in distributed computations \cite{mapreduce,lagrange,Polynomial,entangle,codedsketch,opt-recovery,short}.
In particular, in~\cite{Grad_Dimakis}, a coded computing framework, called \emph{gradient
coding},  is proposed which allows the master node to collect  and aggregate the gradient vectors, with minimum communication cost, while the effect of stragglers is mitigated. In \cite{ye2018communication}, a trade-off between communication cost, computation load, and straggler tolerance is characterized in a homogeneous distributed system. 
The scheme presented in \cite{fahim2019lagrange} extends Short-Dot problem~\cite{short} to calculate a polynomial function over a specific data set and reach a trade-off between the computation load of each worker node and the minimum number of worker nodes needed for the master node to recover the results.
 In stochastic gradient decent (SGD), the algorithm works even with an unbiased approximation of the aggregated gradient vector \cite{bitar2020stochastic}. This motivates researchers to explore approximated versions of  gradient coding~\cite{grad_app_dimakis,glasgow2020approximate,charles2017approximate,wang2019erasurehead,bitar2020stochastic}. In  particular, \cite{grad_app_dimakis,charles2017approximate}, some graph-based analysis is used to propose approximate gradient coding schemes, which  reduce the computation load at the cost of the optimization accuracy. In \cite{wang2019erasurehead}, 
a distributed training method in the presence of stragglers is presented using the approximate gradient coding of \cite{charles2017approximate}, and  some reduction in the total training time is reported. In \cite{wang2019fundamental}, a fundamental trade-off among the computation load, the accuracy of the result, and the number of stragglers is characterized. Furthermore, the authors introduced two schemes to achieve the trade-off.   A near-optimal straggler mitigation scheme is proposed in~\cite{li2018near}, and is extended to heterogeneous cases, where servers have different computation power and  communication capabilities.  In \cite{wang2019heterogeneity}, the idea of \cite{Grad_Dimakis} is extended to propose a  gradient coding scheme in heterogeneous distributed systems. It is shown that the proposed scheme in \cite{wang2019heterogeneity} is optimal in terms of the whole task time, given the accurate estimation of the computation power of each worker node. In the mentioned scheme, the vector sent by each worker node is the size of gradient vectors.

\begin{figure}[htp]
 	
 	\begin{center}
 		\scalebox{0.85}{
 	
 		\tikzset {_g00n00rhh/.code = {\pgfsetadditionalshadetransform{ \pgftransformshift{\pgfpoint{0 bp } { 0 bp }  }  \pgftransformrotate{0 }  \pgftransformscale{2 }  }}}
 		\pgfdeclarehorizontalshading{_ifqmxew5g}{150bp}{rgb(0bp)=(0.65,0.81,0.87);
 			rgb(37.5bp)=(0.65,0.81,0.87);
 			rgb(62.5bp)=(0.14,0.33,0.54);
 			rgb(100bp)=(0.14,0.33,0.54)}
 		\tikzset{every picture/.style={line width=0.75pt}} %set default line width to 0.75pt        
 		
 		\begin{tikzpicture}[x=0.75pt,y=0.75pt,yscale=-1,xscale=1]
 		%uncomment if require: \path (0,368); %set diagram left start at 0, and has height of 368
 		
 		%Image [id:dp867602982094053] 
 		\draw (139.35,251.95) node  {\includegraphics[width=48.53pt,height=29.63pt]{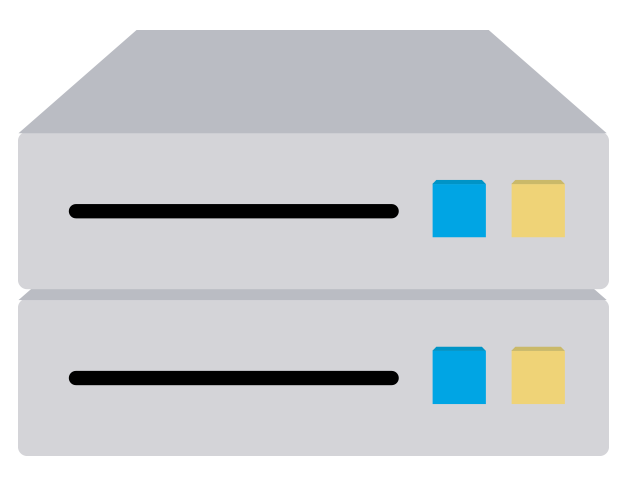}};
 		%Shape: Brace [id:dp5154104840571694] 
 		\draw   (305.9,320.56) .. controls (305.95,325.23) and (308.3,327.54) .. (312.97,327.49) -- (373.1,326.92) .. controls (379.77,326.85) and (383.12,329.15) .. (383.16,333.82) .. controls (383.12,329.15) and (386.43,326.79) .. (393.1,326.72)(390.1,326.75) -- (453.23,326.15) .. controls (457.9,326.1) and (460.21,323.75) .. (460.16,319.08) ;
 		%Shape: Brace [id:dp605095121568233] 
 		\draw   (107.92,320.14) .. controls (107.92,324.81) and (110.25,327.14) .. (114.92,327.14) -- (178.52,327.14) .. controls (185.19,327.14) and (188.52,329.47) .. (188.52,334.14) .. controls (188.52,329.47) and (191.85,327.14) .. (198.52,327.14)(195.52,327.14) -- (262.12,327.14) .. controls (266.79,327.14) and (269.12,324.81) .. (269.12,320.14) ;
 		%Straight Lines [id:da44959073236918057] 
 		\draw    (138.5,141.56) -- (312,140.56) ;
 		%Straight Lines [id:da1185234746228212] 
 		\draw    (238.3,151.36) -- (312,150.56) ;
 		%Straight Lines [id:da29218723229692367] 
 		\draw    (138.96,234.81) -- (138.5,141.56) ;
 		%Straight Lines [id:da018264807859732057] 
 		\draw    (238.72,234.41) -- (238.3,151.36) ;
 		%Straight Lines [id:da22047072186463756] 
 		\draw    (328.96,235.51) -- (328.4,158.4) ;
 		%Straight Lines [id:da7710083693911531] 
 		\draw    (352.5,139.56) -- (526.5,140.56) ;
 		%Straight Lines [id:da25344821714961685] 
 		\draw    (352.9,149.56) -- (429,149) ;
 		%Straight Lines [id:da6106075168676341] 
 		\draw    (429.36,235.42) -- (429,149) ;
 		%Straight Lines [id:da5038943720545286] 
 		\draw    (527.76,235.26) -- (526.5,140.56) ;
 		%Straight Lines [id:da23590998889309578] 
 		\draw    (58.9,131.56) -- (312.9,132.56) ;
 		%Straight Lines [id:da939178218563474] 
 		\draw    (59.36,234.01) -- (58.9,131.56) ;
 		%Straight Lines [id:da14948951187753967] 
 		\draw    (358.9,132.96) -- (596.4,133.24) ;
 		%Straight Lines [id:da925353196818061] 
 		\draw    (598.4,235.24) -- (596.4,133.24) ;
 		%Notched Right Arrow [id:dp4767578470849376] 
 		\path  [shading=_ifqmxew5g,_g00n00rhh] (322.75,94.8) -- (322.67,74.34) -- (315.45,74.36) -- (329.83,60.66) -- (344.32,74.25) -- (337.1,74.28) -- (337.18,94.75) -- (329.94,87.56) -- cycle ; % for fading 
 		\draw   (322.75,94.8) -- (322.67,74.34) -- (315.45,74.36) -- (329.83,60.66) -- (344.32,74.25) -- (337.1,74.28) -- (337.18,94.75) -- (329.94,87.56) -- cycle ; % for border 
 		
 		%Image [id:dp6327291983579788] 
 		\draw (330.7,136.7) node  {\includegraphics[width=51.45pt,height=42.45pt]{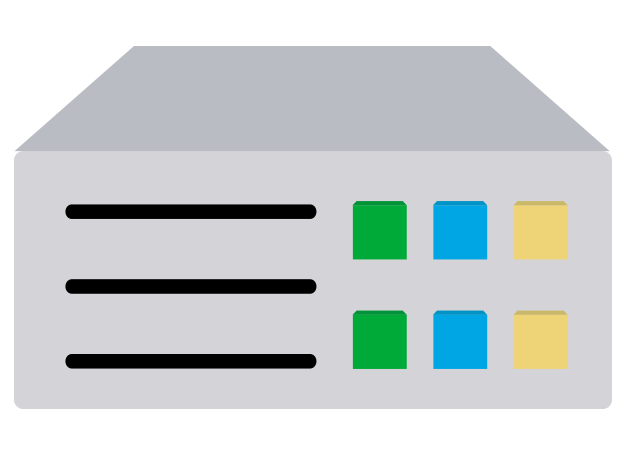}};
 		%Image [id:dp2986236324153704] 
 		\draw (62.85,312.2) node  {\includegraphics[width=21.22pt,height=24.75pt]{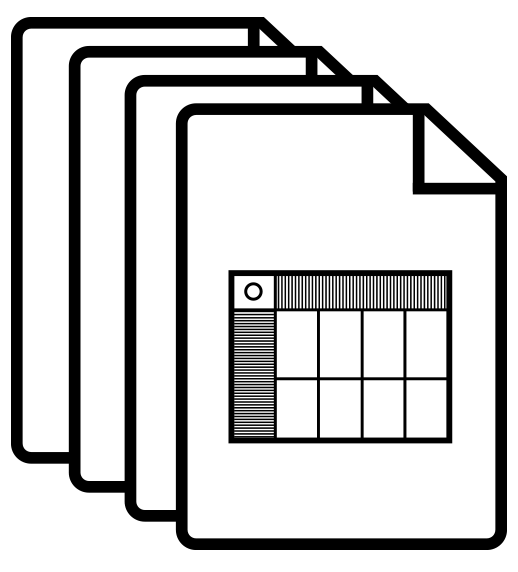}};
 		%Image [id:dp5215930231197152] 
 		\draw (138.85,285.95) node  {\includegraphics[width=17.78pt,height=21.38pt]{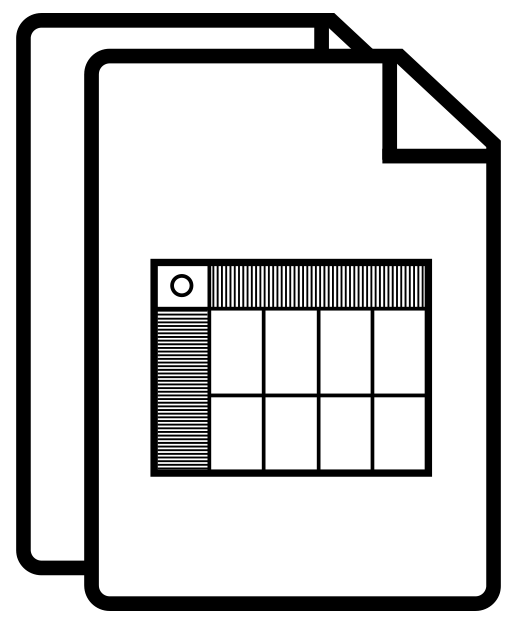}};
 		%Flowchart: Summing Junction [id:dp48744591569381357] 
 		\draw  [color={rgb, 255:red, 208; green, 2; blue, 27 }  ,draw opacity=1 ][line width=4.5]  (150.5,286.18) .. controls (150.5,271.7) and (161.72,259.96) .. (175.55,259.96) .. controls (189.38,259.96) and (200.6,271.7) .. (200.6,286.18) .. controls (200.6,300.66) and (189.38,312.4) .. (175.55,312.4) .. controls (161.72,312.4) and (150.5,300.66) .. (150.5,286.18) -- cycle ; \draw  [color={rgb, 255:red, 208; green, 2; blue, 27 }  ,draw opacity=1 ][line width=4.5]  (157.84,267.64) -- (193.26,304.72) ; \draw  [color={rgb, 255:red, 208; green, 2; blue, 27 }  ,draw opacity=1 ][line width=4.5]  (193.26,267.64) -- (157.84,304.72) ;
 		%Image [id:dp633568925549651] 
 		\draw (239.35,256.2) node  {\includegraphics[width=48.53pt,height=37.5pt]{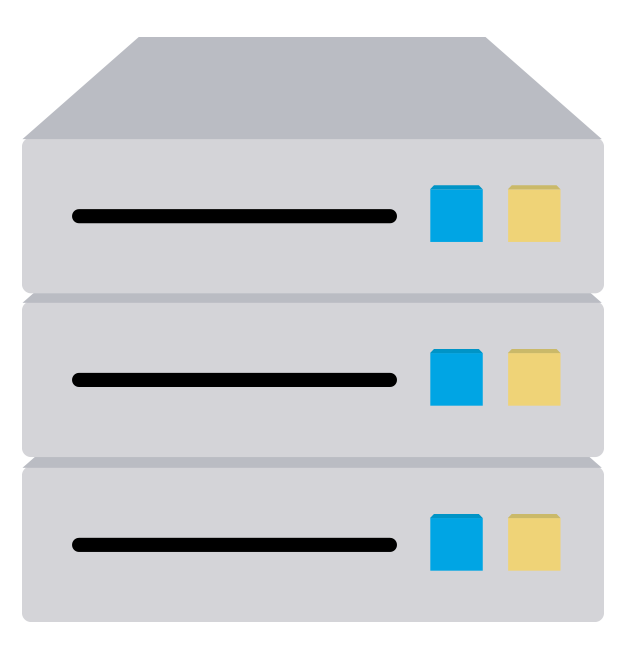}};
 		%Image [id:dp2739181038482721] 
 		\draw (241.35,296.2) node  {\includegraphics[width=20.03pt,height=22.5pt]{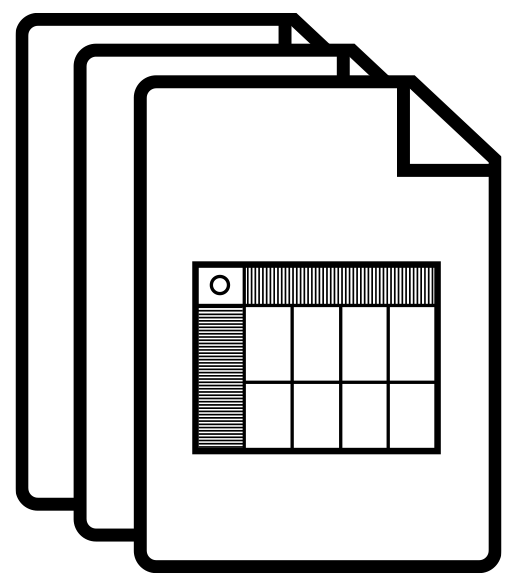}};
 		%Image [id:dp9306813527068525] 
 		\draw (329.54,264.2) node  {\includegraphics[width=48.06pt,height=47.25pt]{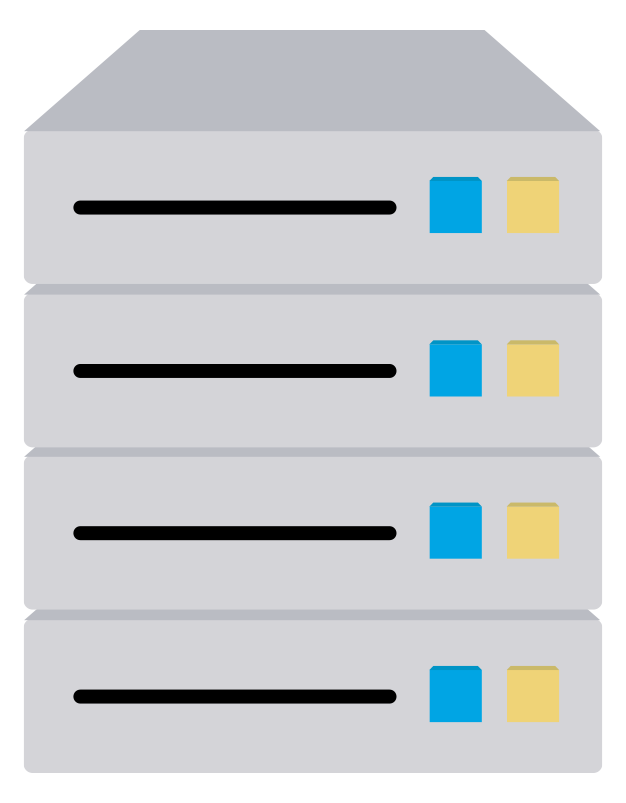}};
 		%Image [id:dp21500201492078497] 
 		\draw (369.76,286.37) node  {\includegraphics[width=45pt,height=41.7pt]{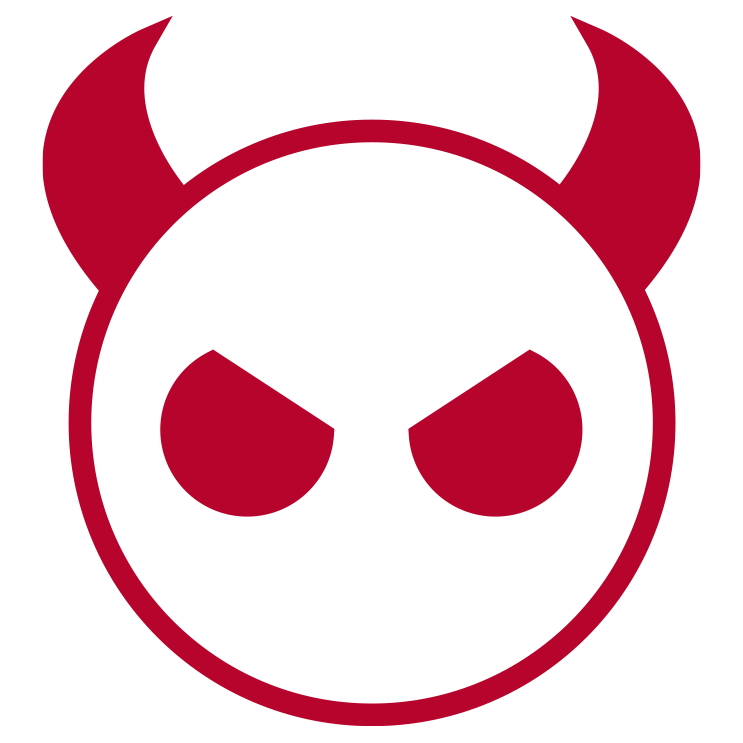}};
 		%Image [id:dp37907132145802036] 
 		\draw (59.54,262.7) node  {\includegraphics[width=48.06pt,height=47.25pt]{server4.png}};
 		%Image [id:dp32706046371203956] 
 		\draw (329.85,310.5) node  {\includegraphics[width=21.22pt,height=24.75pt]{data4.png}};
 		%Image [id:dp8204712722119036] 
 		\draw (429.04,244.45) node  {\includegraphics[width=46.56pt,height=17.63pt]{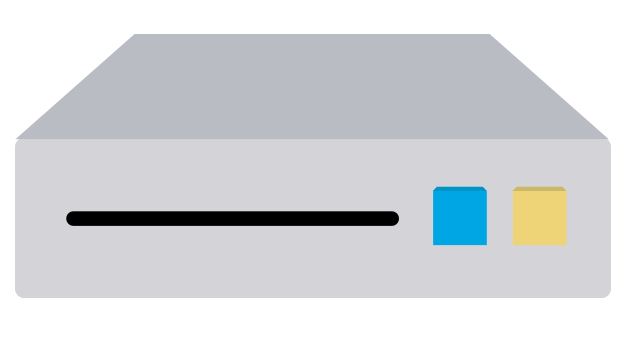}};
 		%Image [id:dp9167416594565334] 
 		\draw (431.29,270.7) node  {\includegraphics[width=14.69pt,height=19.5pt]{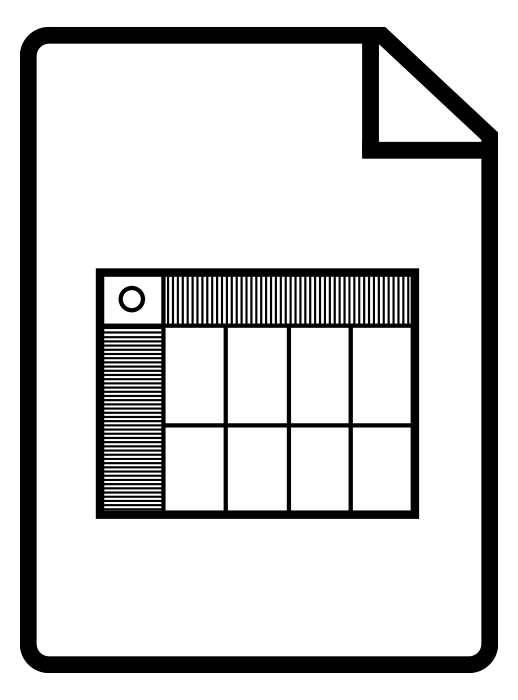}};
 		%Image [id:dp17118633699863994] 
 		\draw (527.85,257.2) node  {\includegraphics[width=48.53pt,height=37.5pt]{server3.png}};
 		%Image [id:dp7654207956797183] 
 		\draw (532.35,300.7) node  {\includegraphics[width=20.03pt,height=22.5pt]{data3.png}};
 		%Image [id:dp902898914654245] 
 		\draw (598.35,252.45) node  {\includegraphics[width=48.53pt,height=29.63pt]{server2.png}};
 		%Image [id:dp2286044500419595] 
 		\draw (601.85,299.45) node  {\includegraphics[width=17.78pt,height=21.38pt]{data2.png}};

 		% Text Node
 		\draw (384.01,218.84) node  [font=\Large] [align=left] {...};
 		% Text Node
 		\draw (299,334) node [anchor=north west][inner sep=0.75pt]  [color={rgb, 255:red, 208; green, 2; blue, 27 }  ,opacity=1 ] [align=left] {$\displaystyle Adversarial\ worker\ nodes$ };
 		% Text Node
 		\draw (290.8,218.44) node  [font=\Large] [align=left] {...};
 		% Text Node
 		\draw (192.61,218.84) node  [font=\Large] [align=left] {...};
 		% Text Node
 		\draw (88,333) node [anchor=north west][inner sep=0.75pt]  [color={rgb, 255:red, 208; green, 2; blue, 27 }  ,opacity=1 ] [align=left] {$\displaystyle \ \ Straggling\ worker\ nodes$ };
 		% Text Node
 		\draw (481.41,217.84) node  [font=\Large] [align=left] {...};
 		% Text Node
 		\draw (94.4,219.24) node  [font=\Large] [align=left] {...};
 		% Text Node
 		\draw (294,99) node [anchor=north west][inner sep=0.75pt]  [font=\footnotesize]  {$Master\ Node$};
 		% Text Node
 		\draw (292,35) node [anchor=north west][inner sep=0.75pt]    {$\mathbf{g} =\sum ^{K}_{k=1}\mathbf{g}_{k}$};
 		% Text Node
 		\draw (35.16,304.65) node [anchor=north west][inner sep=0.75pt]  [font=\tiny]  {$\Gamma _{1}$};
 		% Text Node
 		\draw (113.66,281) node [anchor=north west][inner sep=0.75pt]  [font=\tiny]  {$\Gamma _{i}$};
 		% Text Node
 		\draw (205.06,291.3) node [anchor=north west][inner sep=0.75pt]  [font=\tiny]  {$\Gamma _{i+s}$};
 		% Text Node
 		\draw (302.16,304.4) node [anchor=north west][inner sep=0.75pt]  [font=\tiny]  {$\Gamma _{j}$};
 		% Text Node
 		\draw (395.86,266.8) node [anchor=north west][inner sep=0.75pt]  [font=\tiny]  {$\Gamma _{j+a}$};
 		% Text Node
 		\draw (490.26,295.3) node [anchor=north west][inner sep=0.75pt]  [font=\tiny]  {$\Gamma _{N-1}$};
 		% Text Node
 		\draw (573.96,295.8) node [anchor=north west][inner sep=0.75pt]  [font=\tiny]  {$\Gamma _{N}$};
 		% Text Node
 		\draw (37.6,167.2) node [anchor=north west][inner sep=0.75pt]    {$\tilde{\mathbf{g}}_{1}$};
 		% Text Node
 		\draw (119.8,167.2) node [anchor=north west][inner sep=0.75pt]    {$\tilde{\mathbf{g}}_{i}$};
 		% Text Node
 		\draw (210.8,168) node [anchor=north west][inner sep=0.75pt]    {$\tilde{\mathbf{g}}_{i+s}$};
 		% Text Node
 		\draw (310.16,177.65) node [anchor=north west][inner sep=0.75pt]    {$\tilde{\mathbf{g}}_{j}$};
 		% Text Node
 		\draw (398.2,167.2) node [anchor=north west][inner sep=0.75pt]    {$\tilde{\mathbf{g}}_{j+a}$};
 		% Text Node
 		\draw (494,167.4) node [anchor=north west][inner sep=0.75pt]    {$\tilde{\mathbf{g}}_{N-1}$};
 		% Text Node
 		\draw (576,167.8) node [anchor=north west][inner sep=0.75pt]    {$\tilde{\mathbf{g}}_{N}$};

 		\end{tikzpicture}
 	}
 	\end{center}
 \caption{An overview of a heterogeneous distributed system including $N$ worker nodes, with up to $s$ straggling and $a$ adversarial worker nodes. The objective is to calculate the aggregated gradient vector $\mathbf{g}=\sum_{k=1}^{K}\mathbf{g}_k$, where $\mathbf{g}_k$ is the partial gradient vector over the $k$th partition of data set. Each worker node has a subset of data partitions and encodes the locally computed partial gradients. In other words, $n$th worker node sends $\tilde{\mathbf{g}}_n$ to the master node as a linear combination of its computed partial gradient vectors. }
 \label{fig2}
 \end{figure}

In this paper, we propose an alternative approach for gradient coding, that can \emph{optimally} handle the general heterogeneous gradient coding problem, consisting of a variety of worker nodes with different size of storage and computation power, where some of them are stragglers and some others are  controlled by the adversary. The proposed scheme also allows us to address the cases, where the objective is to calculate the aggregated gradient vector exactly, or approximately. 
%
%which is common in practical scenarios.  
%
%which  not only very simply reproduces the optimum results of XX and ZZZ, but also allows to handle important cases  
%
%
%Here in this paper, we focus on the general heterogeneous gradient coding problem consisting of a variety of worker nodes with different size of storage and computation power, which is common in practical scenarios.
Consider a typical distributed system illustrated in Fig.~\ref{fig2}, consisting of a master node and $N \in \mathbb{N}$ worker nodes, with up to $s \in \mathbb{N}$ stragglers,  and and up to $a\in\mathbb{N}$ adversaries. The adversarial worker nodes may send arbitrary incorrect results to the master node, aiming to make the final result incorrect. Generally, each worker node computes partial gradient vectors/matrices based on the local data set and returns a coded vector/matrix to the master node. The master node receives the results of the fastest worker nodes until it is capable of recovering the final result. 

Our main contribution is characterizing the optimum trade-off among the communication cost, the number of adversarial worker nodes $a$, and straggler tolerance $s$, for any \emph{arbitrary} data placement among worker nodes in heterogeneous distributed systems. The $k$th data partition is allocated to the worker nodes which belong to set $\mathcal{A}_k$, and the minimum communication cost is determined by $\min_{k\in[K]}|\mathcal{A}_k|$.  The proposed gradient coding scheme, which achieves the minimum communication cost, uses the definition of a universal polynomial function developed such that each worker node can compute a distinct point of the function, using only the data sets locally available in the worker node. The results are sent to the master node, which are interpolated to recover the universal function, using error-correcting decoding. This gradient coding scheme can cover not only some existing results in this area  but also can target new problems such as computation of a polynomial function of the aggregated gradient matrix and approximate gradient coding with low complexity decoding and better numerical stability. The proposed approximate gradient coding can cover two cases: (1) when the repetition of the computation in the worker nodes does not support the restriction that we have on the communication cost, and (2) when the number of stragglers appears to be more than what we expected in system design. 

  % We propose a heterogeneous gradient coding scheme that can not only tolerate $s$ stragglers but also can resist against $a$ adversarial worker nodes, as long as $N\ge2a+s+1$. 
%In this setting,  each worker node is allocated a subset of the data set based on its storage and computation power.. We propose a heterogeneous gradient coding scheme that can not only tolerate $s$ stragglers but also can resist against $a$ adversarial worker nodes, as long as $N\ge2a+s+1$. Also, we present a trade-off between communication cost, the number of adversarial worker nodes, straggler tolerance, and the minimum number of replications of the data partition among worker nodes. 

\subsection{Summary of the Contributions:}
Our main contributions for  gradient coding problem in heterogeneous distributed systems are summarized as follows: 
\begin{itemize}
	\item
	We establish a converse bound on the minimum communication cost of the exact gradient coding problem with linear encoding  in the presence of $s$ straggling and $a$ adversarial worker nodes for any arbitrary data partitioning among worker nodes (Section \ref{scheme}). 
	\item
	We propose an achievable gradient coding scheme,  based on introducing a universal polynomial function. Regardless of how the data set is distributed among the worker nodes, the proposed achievable scheme is optimal and meets the converse (Subsection \ref{general}). 
	\item
	By the universal polynomial function, we extend the proposed scheme to another interesting case which is  the matrix polynomial computation of the aggregated gradient matrix.  In this case, the master node aims  recover a function of the aggregated gradient matrix (Section \ref{funcComp}).

	\item Motivated by  machine learning applications, and relaxing the restricting conditions on the minimum number of required computation to recover the exact results, we propose an approximated scheme  which is numerically stable with low computational complexity (Section \ref{appSec}).
\end{itemize}
\subsection{Notation}
In this paper matrices and vectors are denoted by upper boldface letters and lower boldface letters respectively. $\mathcal{C}[a,b]$ denotes the space of all continuous functions on $[a,b]$, where $[a,b]$ is a closed interval. For $n_1, n_2\in\mathbb{Z}$ the notation $[n_1:n_2]$ represents the set $\{n_1,\dots n_2\}$. Also, $[n]$ denotes the set $\{1,\dots,n\}$ for $n\in\mathbb{N}$. 
Furthermore, the cardinality of set $\mathcal{S}$ is denoted by $|\mathcal{S}|$.  $\norm{f}$ denotes the maximum norm of function $f(x)$ over $x$ domain, i.e., $\norm{f}=\max\limits_{x\in[a,b]}|f(x)|$.  The $i$th element of a vector $\mathbf{g}$, is denoted by $\mathbf{g}[i]$.

\section{Problem formulation}
Consider a deep neural network with $L$ layers consisting of $L$ parameter matrices $\mathbf{\Omega}_\ell$ for $\ell=1,\dots L$, which are updated iteratively during the training process. Given a training data set $\mathcal{D}=\{(\mathbf{x}_i,y_i)\}_{i=1}^{M}$ for some integer $M$, where $\mathbf{x}_i$ is the $i$th input sample vector and $y_i$ is the corresponding label. The goal is to learn the parameter matrices $\mathbf{\Omega}_\ell$ for $\ell=1\dots L$ by minimizing the following empirical  loss function
\begin{align}
J(\mathcal{D};\mathbf{\Omega}_1,\dots \mathbf{\Omega}_L)=\frac{1}{|\mathcal{D}|}\sum_{(\mathbf{x}_i,y_i)\in \mathcal{D}}{J(\mathbf{x}_i,y_i;\mathbf{\Omega}_1,\dots \mathbf{\Omega}_L)}.
\end{align}
More specifically, this optimization problem is usually solved by gradient descent algorithm, which starts with some initial value $\mathbf{\Omega}^{(0)}_\ell$, and then in each iteration $t$ updates this parameter matrices as follows
\begin{align}\label{update}
\bm{\omega}_\ell^{(t+1)}(k)=\bm{\omega}_\ell^{(t)}(k)-\eta\nabla_{\bm{\omega}_\ell^{(t)}(k)}{J(\mathcal{D};\mathbf{\Omega}_1^{(t)},\dots \mathbf{\Omega}_L^{(t)})},
\end{align}
where $\eta\in\mathbb{R}$ is the learning rate, $\bm{\omega}_{\ell}^{(t)}(k)$ is the weight of the $k$th neuron of the $\ell$th layer at the $t$th iteration, and $\mathbf{g}^{(t,\ell)}\triangleq\nabla_{\bm{\omega}_\ell^{(t)}(k)}{J(\mathcal{D};\mathbf{\Omega}_1^{(t)},\dots \mathbf{\Omega}_L^{(t)})}$ is the gradient of the loss function at the current parameters of layer $\ell$ and iteration $t$ over whole data set.  For simplicity of presentation, in the rest of the paper we denote $\mathbf{g}^{(t,\ell)}$ by $\mathbf{g}$. 
Now consider a heterogeneous distributed system with one master node and $N$ worker nodes ${W}_1,\dots,{W}_N$ consisting of up to $s\in\mathbb{N}$ stragglers and up to $a\in\mathbb{N}$ adversarial worker nodes. The training data set $\mathcal{D}$ is partitioned into $K$  equal-size subsets $\mathcal{D}=\{\mathcal{D}_1,\dots\mathcal{D}_{K}\}$.
 The main task of this distributed system is to collaboratively compute the aggregated gradient vector $\mathbf{g}=\sum\limits_{k=1}^K{\mathbf{g}_k}$, where $\mathbf{g}_k\in\mathbb{R}^d$ is the partial
gradient over $\mathcal{D}_k$ for $k\in[K]$. We have an arbitrary data placement among worker nodes. In other words, due to
the limited computing power of each worker node, a subset of data partitions denoted by $\Gamma_n$ with the size of $0<|\Gamma_n|\le K$, is allocated to worker node $W_n$, where $\Gamma_n\subseteq\mathcal{D}$ for $n\in[N]$. Let  $\mathcal{A}_k$ denote the subset of worker nodes, where $\mathcal{D}_k$ is assigned to each of them. In other words,
\begin{align}
\mathcal{A}_k=\{ {W}_n: \mathcal{D}_k\in\Gamma_n, \text{for }  n\in[N]\},
\end{align}
 where $k\in[K]$. 
Also, we define 
\begin{align}
	r=\min\limits_{k\in[K]}|\mathcal{A}_k|,
\end{align}
 as the minimum number of replications of the data partitions. 
Generally, the $n$th worker node encodes  the locally computed partial gradients using a \emph{linear} encoding function
\begin{align}\label{encoding1}
	\mathcal{E}_n:(\mathbb{R}^{d})^{|\Gamma_n|}\to \mathbb{R}^{C},
\end{align}
where we define the dimension of the vector sent by each worker node, i.e., $C$, as the communication cost. This encoding function maps the raw partial gradients  to the coded partial gradient vector $\tilde{\mathbf{g}}_n\in\mathbb{R}^C$  for $n$th worker node as follows
\begin{align}
	 \tilde{\mathbf{g}}_n=\mathcal{E}_n\big( \{\mathbf{g}_i\}_{i: \mathcal{D}_i\in\Gamma_n}\big),
\end{align}
which is sent to the master node. The master node waits for the results from a set of fastest worker nodes, denoted by $\mathcal{F}$. Then it recovers the desired aggregated gradient vector as follows
\begin{align}
 \mathbf{g}=\mathcal{H}\big(\mathcal{F},\{\tilde{\mathbf{g}}_j\}_{j\in\mathcal{F}}\big),
\end{align}
where
\begin{align}
\mathcal{H}:\mathcal{F}\times(\mathbb{R}^{C})^{|\mathcal{F}|}\to \mathbb{R}^{d},
\end{align}  
is a decoding function in the proposed scheme. Generally, the resulting distributed system needs to tolerate the presence of $s$ straggling and $a$ adversarial worker nodes  and the master node does not know in advance which worker nodes are stragglers or adversaries. The adversarial worker node may send arbitrary incorrect results to the master node. Thus, the master node receives $|\mathcal{F}|=N-s$ answers from non-straggling worker nodes in which at most $a$ of them are adversarial.  The proposed exact gradient coding scheme is determined by the pair $(\mathcal{E}_{n\in\mathbb{N}},\mathcal{H})$.
A communication cost $C$ is achievable if there exists a gradient coding scheme in a way that the $n$th worker node locally encodes its computed partial gradient vectors by designing the encoding function $\mathcal{E}_n$ so that from the answers of any $N-s$ worker nodes, the master node can recover final result using the designed decoding function $\mathcal{H}$.  The goal is to find the optimal gradient coding scheme for arbitrary data placements with linear encoding functions which leads to the minimum communication cost. The minimum communication cost is the minimum value of all achievable communication costs and is denoted by $C^*(s,a,\{\mathcal{A}_k\}_{k=1}^K)$.

This problem formulation is for exact gradient coding scheme. In Section~\ref{funcComp} and Section~\ref{appSec}, we change this formulation to cover the polynomial computation of the aggregated gradient matrix and the calculation of the approximate aggregated gradient vector respectively.

\section{Exact Gradient Coding Scheme}\label{scheme}
In this section, we introduce the proposed gradient coding scheme. We first propose a converse bound on the minimum communication cost in the following theorem. Then, we describe the achievable scheme.
\subsection{Main Result}
\begin{theorem}\label{th1}
	In an exact gradient coding scheme with $s$ straggling and $a$ adversarial worker nodes, and data assignments $\{\mathcal{A}_k\}_{k=1}^K$, the minimum communication cost is characterized by
	\begin{align}\label{comm.cost}
	C^*(s,a,\{\mathcal{A}_k\}_{k=1}^K)=\frac{d}{\min\limits_{k\in[K]}|\mathcal{A}_k|-2a-s},
	\end{align}
	where $d$ is the dimension of the gradient vectors.
\end{theorem}

\begin{remark}
The proof of converse can be found in Subsection \ref{converse}, where we show that for linear encoding functions, if the communication cost is less than $C^*$, then in the presence of $s$ straggling and $a$ adversarial worker nodes we can find scenarios such that for two different sets of partial gradient vectors the decoder in the master node will receive the same inputs. Thus, it can not have different outputs.
\end{remark}
\begin{remark}
		The achievable scheme can be found in Subsection \ref{general}. It is based on the definition of \emph{a universal polynomial function} developed such that each worker node can compute a distinct point of that function, using only its local data set, and the master node can interpolate the universal polynomial function from samples received from the workers,  using error-correcting Reed-Solomon decoding methods. This polynomial function is such that its values at some specific points are equal to the partitions of the aggregated gradient vector.
\end{remark}
\begin{remark}
	In \eqref{comm.cost}, the minimum communication cost is determined by $\min\limits_{k\in[K]}|\mathcal{A}_k|$, irrespective of the structure of $\mathcal{A}_k$ or their connections. This allows us to have resource allocation algorithms in gradient coding problems based on the computation power of the worker nodes.
\end{remark}
\begin{remark}
	There is no gain in having non-uniform redundancy of data partitions in terms of reducing the communication cost. Therefore, equal replication for all data partitions would be a better choice. Note that having uniform redundancy of data partitions is not equivalent to having the uniform computation load in the worker nodes. In other words, we can have uniform redundancy in job assignment while the computation loads of worker nodes are different. Equation \eqref{comm.cost} can be used to design optimum job assignment subject to the computation power of each worker node.   
\end{remark}
\begin{remark}
	Consider a particular case where the minimum communication cost is the bottleneck of a distributed system and cannot be greater than $R$, for some $R\le	C^*(s,a,\{\mathcal{A}_k\}_{k=1}^K)$. Theorem \ref{th1} suggests that we can sacrifice the accuracy of the final result in order to reduce the communication cost.
	Without loss of generality assume that $r=|\mathcal{A}_1|\le|\mathcal{A}_2|\le\dots\le|\mathcal{A}_K|$.  Let $j$ be the smallest integer such that
	\begin{align}
	R\ge \frac{d}{|\mathcal{A}_{j}|-2a-s}.
	\end{align} 
	Then we can design a scheme such that the master node can recover the value of $\sum_{k=j}^K\mathbf{g}_k$ accurately instead of $\sum_{k=1}^K\mathbf{g}_k$. In Section \ref{appSec}, we present another approach for approximate gradient coding for the case where $R\le	C^*(s,a,\{\mathcal{A}_k\}_{k=1}^K)$.
\end{remark}
\begin{remark}
Assume that $r=|\mathcal{A}_1|\le|\mathcal{A}_2|\le\dots\le|\mathcal{A}_K|$. Consider another particular case where $r\le 2a+s$. Theorem \ref{th1} implies that the master node can compute the value of $\sum_{k=\ell}^{K}\mathbf{g}_k$, where $\ell$ is the smallest integer such that $|\mathcal{A}_{\ell}|>2a+s$. 
\end{remark}
\subsection{Motivating Example}
Before going into details of the proposed scheme, we demonstrate the main idea of our scheme through a simple example. Then we generalize this approach in the next subsection. In \figref{fig1}, a master node and five worker nodes are shown, where one of the worker nodes is straggler and assume $a=0$. The training data set $\mathcal{D}$ is partitioned into 5 equal-size subsets $\mathcal{D}=\{\mathcal{D}_1,\dots\mathcal{D}_{5}\}$, where each of them is assigned to at least $r=3$ worker nodes.
Each worker node has a subset of data partitions based on its limited storage and computation power. Also, worker nodes have the current parameters of the model sent by the master node. Thus, each worker node computes its partial gradients at the $t$th iteration. 
\tikzset{every picture/.style={line width=0.75pt}} %set default line width to 0.75pt        
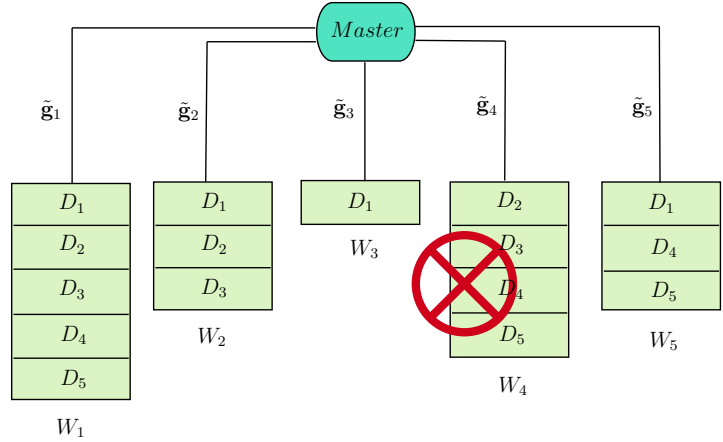
\begin{figure}[htp]
	
	\begin{center}
		\scalebox{0.7}{%

\tikzset{every picture/.style={line width=0.75pt}} %set default line width to 0.75pt        

\begin{tikzpicture}[x=0.75pt,y=0.75pt,yscale=-1,xscale=1]
%uncomment if require: \path (0,400); %set diagram left start at 0, and has height of 400

%Shape: Rectangle [id:dp6078739544873475] 
\draw  [fill={rgb, 255:red, 184; green, 233; blue, 134 }  ,fill opacity=0.49 ] (525.09,164.66) -- (610.46,164.66) -- (610.46,256.8) -- (525.09,256.8) -- cycle ;
%Shape: Rectangle [id:dp6022632709271587] 
\draw  [fill={rgb, 255:red, 184; green, 233; blue, 134 }  ,fill opacity=0.49 ] (99.81,165.71) -- (185.18,165.71) -- (185.18,321.47) -- (99.81,321.47) -- cycle ;
%Straight Lines [id:da8643108697775659] 
\draw    (101.13,195.92) -- (184.17,195.92) ;
%Straight Lines [id:da24989974394524817] 
\draw    (101.13,227.49) -- (184.17,227.49) ;
%Straight Lines [id:da9805583936168607] 
\draw    (101.13,291.69) -- (184.17,291.69) ;
%Straight Lines [id:da9158617137707543] 
\draw    (101.13,260.11) -- (184.17,260.11) ;
%Shape: Rectangle [id:dp3829239758020242] 
\draw  [fill={rgb, 255:red, 184; green, 233; blue, 134 }  ,fill opacity=0.49 ] (202.09,164.66) -- (287.46,164.66) -- (287.46,256.8) -- (202.09,256.8) -- cycle ;
%Straight Lines [id:da26399075859766974] 
\draw    (203.41,195.92) -- (286.44,195.92) ;
%Straight Lines [id:da4499646603997036] 
\draw    (203.41,226.44) -- (286.44,226.44) ;
%Shape: Rectangle [id:dp2552771800697273] 
\draw  [fill={rgb, 255:red, 184; green, 233; blue, 134 }  ,fill opacity=0.49 ] (308.42,163.61) -- (393.78,163.61) -- (393.78,195.18) -- (308.42,195.18) -- cycle ;
%Shape: Rectangle [id:dp3808257038106315] 
\draw  [fill={rgb, 255:red, 184; green, 233; blue, 134 }  ,fill opacity=0.5 ] (415.76,164.66) -- (501.12,164.66) -- (501.12,290.95) -- (415.76,290.95) -- cycle ;
%Straight Lines [id:da8454493845348592] 
\draw    (417.07,195.92) -- (500.11,195.92) ;
%Straight Lines [id:da05840727308724203] 
\draw    (417.07,226.44) -- (500.11,226.44) ;
%Straight Lines [id:da07697323882350493] 
\draw    (417.07,259.06) -- (500.11,259.06) ;
%Straight Lines [id:da7494708505160883] 
\draw    (526.46,228.92) -- (609.5,228.92) ;
%Flowchart: Terminator [id:dp5830849585397375] 
\draw  [color={rgb, 255:red, 0; green, 0; blue, 0 }  ,draw opacity=1 ][fill={rgb, 255:red, 80; green, 227; blue, 194 }  ,fill opacity=1 ] (330.9,35.32) -- (379.1,35.32) .. controls (385.36,35.32) and (390.44,44.74) .. (390.44,56.36) .. controls (390.44,67.99) and (385.36,77.41) .. (379.1,77.41) -- (330.9,77.41) .. controls (324.63,77.41) and (319.56,67.99) .. (319.56,56.36) .. controls (319.56,44.74) and (324.63,35.32) .. (330.9,35.32) -- cycle ;
%Straight Lines [id:da31290993605157946] 
\draw    (318.54,53.52) -- (143.36,53.52) ;
%Straight Lines [id:da4100673103740331] 
\draw    (143.36,53.52) -- (143.36,166.13) ;
%Straight Lines [id:da18133360645059993] 
\draw    (319.05,63.52) -- (240.6,63.5) ;
%Straight Lines [id:da5554274179733945] 
\draw    (240.6,63.5) -- (239.6,164.5) ;
%Straight Lines [id:da3851999268078794] 
\draw    (354.29,78.31) -- (354.29,164.08) ;
%Straight Lines [id:da9588220563273919] 
\draw    (455.55,63.05) -- (390.74,62.52) ;
%Straight Lines [id:da1642199412874612] 
\draw    (566.44,52.52) -- (390.24,52.52) ;
%Straight Lines [id:da10368688289586792] 
\draw    (455.55,63.05) -- (455.05,163.55) ;
%Straight Lines [id:da41526582227160613] 
\draw    (566.44,52.52) -- (566.94,165.13) ;
%Shape: Rectangle [id:dp24682257201959756] 
\draw  [color={rgb, 255:red, 255; green, 255; blue, 255 }  ,draw opacity=1 ] (34.6,8.9) -- (675.6,8.9) -- (675.6,378.3) -- (34.6,378.3) -- cycle ;
%Flowchart: Summing Junction [id:dp6142985823280338] 
\draw  [color={rgb, 255:red, 208; green, 2; blue, 27 }  ,draw opacity=1 ][line width=4.5]  (391,238) .. controls (391,218.67) and (406.67,203) .. (426,203) .. controls (445.33,203) and (461,218.67) .. (461,238) .. controls (461,257.33) and (445.33,273) .. (426,273) .. controls (406.67,273) and (391,257.33) .. (391,238) -- cycle ; \draw  [color={rgb, 255:red, 208; green, 2; blue, 27 }  ,draw opacity=1 ][line width=4.5]  (401.25,213.25) -- (450.75,262.75) ; \draw  [color={rgb, 255:red, 208; green, 2; blue, 27 }  ,draw opacity=1 ][line width=4.5]  (450.75,213.25) -- (401.25,262.75) ;
%Straight Lines [id:da7402914276060573] 
\draw    (526.46,195.92) -- (609.5,195.92) ;

% Text Node
\draw (132.36,171.95) node [anchor=north west][inner sep=0.75pt]    {$D_{1}$};
% Text Node
\draw (132.36,201.42) node [anchor=north west][inner sep=0.75pt]    {$D_{2}$};
% Text Node
\draw (132.36,234.04) node [anchor=north west][inner sep=0.75pt]    {$D_{3}$};
% Text Node
\draw (133.38,266.67) node [anchor=north west][inner sep=0.75pt]    {$D_{4}$};
% Text Node
\draw (133.38,298.24) node [anchor=north west][inner sep=0.75pt]    {$D_{5}$};
% Text Node
\draw (130.45,333.87) node [anchor=north west][inner sep=0.75pt]    {$W_{1}$};
% Text Node
\draw (234.64,170.9) node [anchor=north west][inner sep=0.75pt]    {$D_{1}$};
% Text Node
\draw (234.64,201.42) node [anchor=north west][inner sep=0.75pt]    {$D_{2}$};
% Text Node
\draw (234.64,232.99) node [anchor=north west][inner sep=0.75pt]    {$D_{3}$};
% Text Node
\draw (232.11,268.18) node [anchor=north west][inner sep=0.75pt]    {$W_{2}$};
% Text Node
\draw (339.95,170.9) node [anchor=north west][inner sep=0.75pt]    {$D_{1}$};
% Text Node
\draw (342.02,204.84) node [anchor=north west][inner sep=0.75pt]    {$W_{3}$};
% Text Node
\draw (447.29,170.25) node [anchor=north west][inner sep=0.75pt]    {$D_{2}$};
% Text Node
\draw (448.31,201.42) node [anchor=north west][inner sep=0.75pt]    {$D_{3}$};
% Text Node
\draw (448.31,234.04) node [anchor=north west][inner sep=0.75pt]    {$D_{4}$};
% Text Node
\draw (449.32,266.67) node [anchor=north west][inner sep=0.75pt]    {$D_{5}$};
% Text Node
\draw (449.09,303.34) node [anchor=north west][inner sep=0.75pt]    {$W_{4}$};
% Text Node
\draw (558.7,203.9) node [anchor=north west][inner sep=0.75pt]    {$D_{4}$};
% Text Node
\draw (558.71,235.37) node [anchor=north west][inner sep=0.75pt]    {$D_{5}$};
% Text Node
\draw (557.77,271.25) node [anchor=north west][inner sep=0.75pt]    {$W_{5}$};
% Text Node
\draw (218.3,105.4) node [anchor=north west][inner sep=0.75pt]    {$\tilde{\mathbf{g}}_{2}$};
% Text Node
\draw (330.3,103.4) node [anchor=north west][inner sep=0.75pt]    {$\tilde{\mathbf{g}}_{3}$};
% Text Node
\draw (433.3,101.4) node [anchor=north west][inner sep=0.75pt]    {$\tilde{\mathbf{g}}_{4}$};
% Text Node
\draw (545.3,102.4) node [anchor=north west][inner sep=0.75pt]    {$\tilde{\mathbf{g}}_{5}$};
% Text Node
\draw (119.3,104.4) node [anchor=north west][inner sep=0.75pt]    {$\tilde{\mathbf{g}}_{1}$};
% Text Node
\draw (328,47.4) node [anchor=north west][inner sep=0.75pt]    {$Master$};
% Text Node
\draw (556.95,171.9) node [anchor=north west][inner sep=0.75pt]    {$D_{1}$};

\end{tikzpicture}	
		}
	\end{center}
	
	\caption{A distributed system with $N=5$ worker nodes, where up to $s=1$ of them is straggler. The data assignment is $\mathcal{A}_1=\{W_1,W_2,W_3,W_5\}, \mathcal{A}_2=\{W_1,W_2,W_4\}, \mathcal{A}_3=\{W_1,W_2,W_4\}, \mathcal{A}_4=\{W_1,W_4,W_5\}$ and $\mathcal{A}_5=\{W_1,W_4,W_5\}$. Each worker node sends a vector with size of $d/2$, and the master node can recover the aggregated gradient vector $\mathbf{g}=\sum_{k=1}^{5}\mathbf{g}_k$ from the results of any 4 worker nodes. }
	\label{fig1}
\end{figure}

Let us assume each partial gradient can be partitioned into two parts, i.e., $\mathbf{g}_k=\big(\mathbf{g}_k[1],\mathbf{g}_k[2]\big)\in\mathbb{R}^d$. Also,
$\alpha_1,\dots,\alpha_5$  are five distinct real numbers, where $\alpha_n$ is associated to worker node $W_n$ for $n\in[5]$. In this example, for simplicity assume $\alpha_n=n$ for $n\in[5]$. Each worker node sends a linear combination of the elements of its computed partial gradient vectors without knowing the straggler as follows
\begin{align*}
\tilde{\mathbf{g}}_1&=\frac{3}{2}\mathbf{g}_1[1]+\frac{16}{15}(\mathbf{g}_2[1]+\mathbf{g}_3[1])+\frac{2}{3}(\mathbf{g}_4[1]+\mathbf{g}_5[1])-\frac{3}{5}\mathbf{g}_1[2]-\frac{1}{3}(\mathbf{g}_2[2]+\mathbf{g}_3[2])-\frac{1}{6}(\mathbf{g}_4[2]+\mathbf{g}_5[2]),\\
\tilde{\mathbf{g}}_2&=\frac{3}{2}\mathbf{g}_1[1]+\hspace{-1mm}\frac{3}{5}(\mathbf{g}_2[1]+\hspace{-1mm}\mathbf{g}_3[1])-\hspace{-1mm}
\frac{4}{5}\mathbf{g}_1[2]-\hspace{-1mm}\frac{1}{4}(\mathbf{g}_2[2]+\hspace{-1mm}\mathbf{g}_3[2]),\\
\tilde{\mathbf{g}}_3&=\mathbf{g}_1[1]-
\frac{3}{5}\mathbf{g}_1[2],\\
\tilde{\mathbf{g}}_4&=-\frac{1}{3}(\mathbf{g}_2[1]+\mathbf{g}_3[1])+\frac{5}{3}(\mathbf{g}_4[1]+\mathbf{g}_5[1])+\frac{1}{6}(\mathbf{g}_2[2]+\mathbf{g}_3[2])-\frac{2}{3}(\mathbf{g}_4[2]+\mathbf{g}_5[2]),\\
\tilde{\mathbf{g}}_5&=-\frac{3}{2}\mathbf{g}_1[1]+6(\mathbf{g}_4[1]+\mathbf{g}_5[1])+\mathbf{g}_1[2]-
\frac{5}{2}(\mathbf{g}_4[2]+\mathbf{g}_5[2]).
\end{align*}
Note that each worker only transmits one vector with size of $d/2$ to the master node. Consider polynomial function ${f}(x)$ of degree 3 as follows
%\begin{align*}
%	{f}(x)&&=\bigg(&\mathbf{g}_1(0)(\frac{x-4}{4})(\frac{x-5}{5})+\mathbf{g}_2(0)(\frac{x-3}{3})(\frac{x-5}{5})+\\&& &\mathbf{g}_3(0)(\frac{x-3}{3})(\frac{x-5}{5})+\mathbf{g}_4(0)(\frac{x-2}{2})(\frac{x-3}{3})+\\&& &
%	\mathbf{g}_5(0)(\frac{x-2}{2})(\frac{x-3}{3})\bigg)({x+1})+\\&&\bigg(&\mathbf{g}_1[1](\frac{x-4}{5})(\frac{x-5}{6})+\mathbf{g}_2[1](\frac{x-3}{4})(\frac{x-5}{6})+\\&& &\mathbf{g}_3[1](\frac{x-3}{4})(\frac{x-5}{6})+\mathbf{g}_4[1](\frac{x-2}{3})(\frac{x-3}{4})+\\&& &
%	\mathbf{g}_5[1](\frac{x-2}{3})(\frac{x-3}{4})\bigg)({-x}).
%\end{align*}
\begin{align*}
{f}(x)=\bigg(&\mathbf{g}_1[1](\frac{x-4}{-4})+\mathbf{g}_2[1](\frac{x-3}{3})(\frac{x-5}{5})+\mathbf{g}_3[1](\frac{x-3}{3})(\frac{x-5}{5})+\mathbf{g}_4[1](\frac{x-2}{2})(\frac{x-3}{3})\\&+\mathbf{g}_5[1](\frac{x-2}{2})(\frac{x-3}{3})\bigg)({x+1})+\bigg(\mathbf{g}_1[2](\frac{x-4}{-5})+\mathbf{g}_2[2](\frac{x-3}{4})(\frac{x-5}{6}) \\&+ \mathbf{g}_3[2](\frac{x-3}{4})(\frac{x-5}{6})+\mathbf{g}_4[2](\frac{x-2}{3})(\frac{x-3}{4})+
\mathbf{g}_5[2](\frac{x-2}{3})(\frac{x-3}{4})\bigg)({-x}).
\end{align*}
 This polynomial has two important properties: (1) One can verify that $\tilde{\mathbf{g}}_k$ sent by worker node $W_k$ is indeed equal to $f(\alpha_k)$, where $\alpha_k=k$ in this example. Since $f(x)$ is a polynomial of degree $3$, from the result of any 4 worker nodes, the master node can recover $f(x)$. (2) One can observe that $f(0)$ is indeed equal to $\mathbf{g}[1]=\sum_{k=1}^{5}\mathbf{g}_k[1]$, and $f(-1)$ is equal to $\mathbf{g}[2]=\sum_{k=1}^{5}\mathbf{g}_k[2]$. Note that the communication cost of each worker node achieves the optimum value of \eqref{comm.cost}.
\subsection{General Scheme}\label{general}
 To design the gradient coding scheme, we restrict our attention to find a universal polynomial function which is defined as follows. 
% 
% , thus each partition of data set is replicated in at least $s+m$ worker nodes, where $m$ is the communication cost reduction factor \cite{ye2018communication}. Note that 
 \begin{definition}\label{def1}
  Let us partition each partial gradient vector $\mathbf{g}_k$ into $m=r-2a-s$ parts of equal size as $\mathbf{g}_k=~\big(\mathbf{g}_k[1],\dots,\mathbf{g}_k[m]\big)\in\mathbb{R}^d$ for $k\in[K]$. Let $\alpha_1,\alpha_2,\dots,\alpha_N\in\mathbb{R}$ and $\beta_1,\beta_2,\dots,\beta_m\in\mathbb{R}$ be $N+m$ distinct values, where $\beta_{\ell}\ne \alpha_i$ for all $i\in[N]$ and $\ell\in[m]$. We call $f(x)$ a \emph{universal polynomial function} if it has the following properties:
%   The problem of communication and computation efficient
% 	gradient coding using a heterogeneous distributed system in presence of $s$ stragglers and $a$ adversarial worker nodes is easily solvable and we can recover the aggregated gradient vector  if there exists a \textit{universal polynomial function} $f(x)$ with three important properties:
 	\begin{enumerate}
 		\item
 		$f(x)$ is a polynomial function of degree $N-s-1$.
 		\item
   $f(\alpha_j)$ is only a function of partial gradient vectors $\{\mathbf{g}_i\}_{i: \mathcal{D}_i\in\Gamma_j}$ for $j\in[N]$.
 		\item
 	 $f(\beta_{\ell})=\sum\limits_{k=1}^{K}{\mathbf{g}_k}[\ell]$. 
 	\end{enumerate}
 If we have such a function, in the encoding step, the $n$th worker node computes the partial gradients
 over the data partitions $\Gamma_n$ in hand and then sends $f(\alpha_n)$ as a linear combination of its calculated partial gradients. Having received the computed results from $N-s$ worker nodes, the master node is able to recover $f(x)$ using Lagrange interpolation. Then each part of aggregated gradient $\mathbf{g}=\sum\limits_{k=1}^{K}\mathbf{g}_k$, i.e., $\mathbf{g}[\ell]$ can be calculated by evaluation of $f(x)$ at $m$ distinct points $\beta_{\ell}$ for $\ell\in[m]$.   
 \end{definition}

\begin{claim}
	There exists a universal polynomial function in which the properties of Definition~\ref{def1} are satisfied.
	\begin{proof}
	Let us assume $\mathcal{A}_k$ is a subset of worker nodes, where $\mathcal{D}_k$ is assigned to each of them. In other words, $\mathcal{A}_k=\{ {W}_n: \mathcal{D}_k\in\Gamma_n, \text{for }  n\in[N]\}$, where $k\in[K]$. Also, we define $r=\min\limits_{k\in[K]}|\mathcal{A}_k|$ as the minimum number of replications of the data partitions. 
	Assume $N$ distinct values $\alpha_1,\dots,\alpha_N\in\mathbb{R}$, and $m$ other distinct values $\beta_1,\dots,\beta_m\in\mathbb{R}$, where $\alpha_j\ne\beta_{\ell}$ for $\ell\in[m]$ and $j\in[N]$.
				Then, the following function
	\begin{align}
{p}_{\ell}(x)=\sum_{i=1}^{K}{\mathbf{g}_i[\ell]\prod\limits_{j=1\atop j:\mathcal{D}_i\notin\Gamma_j\hfill}^{N}{\frac{x-\alpha_j}{\beta_{\ell}- \alpha_j}}},
\end{align}
is a polynomial function of degree $N-r$ 
		such that $p_{\ell}(\alpha_n)$ is a linear combination of $\{\mathbf{g}_i\}_{i: \mathcal{D}_i\in\Gamma_n}$, for $n\in[N]$. Also, we have $p_{\ell}(\beta_{\ell})=\sum_{i=1}^{K}\mathbf{g}_i[\ell]$. Therefore, the $\ell$th partition of aggregated gradient vector $\mathbf{g}=\sum_{k=1}^{K}$$\mathbf{g}_k$, i.e., $\mathbf{g}[\ell]$ is equal to $p_{\ell}(\beta_{\ell})$. Then, we define function $f(x)$ as
		\begin{align}\label{universal}
			f(x)=\sum_{\ell=1}^{m}\bigg(p_{\ell}(x)\prod\limits_{u=1\hfill\atop u\ne\ell\hfill}^{m}{\frac{x-\beta_u}{\beta_{\ell}-\beta_u }}\bigg).
		\end{align} 
	We can observe that $f(x)$ is the universal polynomial function of degree $N-s-1$, where $f(\alpha_n)$ is a linear combination of the partial gradient vectors $\{\mathbf{g}_i\}_{i: \mathcal{D}_i\in\Gamma_n}$, for $n\in[N]$, and  $\mathbf{g}=(f(\beta_1),\dots,f(\beta_m))$.
	\end{proof} 
\end{claim}
 Let $m=r-2a-s$. Based on the universal polynomial function \eqref{universal}, the exact gradient coding scheme is proposed as follows.
 \begin{enumerate}
 	\item {Encoding Phase at the Worker Nodes:} Each partial gradient vector is partitioned into $m$ groups of equal size, i.e., $\mathbf{g}_k=\big(\mathbf{g}_k[1],\dots,\mathbf{g}_k[m]\big)$ for $k\in[K]$. Also, $\alpha_1,\dots,\alpha_N\in\mathbb{R}$ are $N$ distinct real values, where $\alpha_n$ is assigned to the worker node $W_n$ for $n\in[N]$. In addition, other $m$ distinct real values $\beta_1,\dots,\beta_{m}\in\mathbb{R}$ are chosen.
 	 Then, the $n$th worker node sends a linear combination of the $m$ partitions of its locally computed partial gradient vectors as follows
 	 \begin{align}\label{encoding}
 	\tilde{\mathbf{g}}_n=\sum_{\ell=1}^{m}\bigg(\sum_{i=1\atop i: \mathcal{D}_i\in\Gamma_n}^{K}\hspace{-2mm}{\mathbf{g}_i[\ell]\hspace{-2mm}\prod\limits_{j=1\atop j:\mathcal{D}_i\notin\Gamma_j\hfill}^{N}{\frac{\alpha_n-\alpha_j}{\beta_{\ell}- \alpha_j}}}\bigg)\prod\limits_{u=1\hfill\atop u\ne\ell\hfill}^{m}{\frac{\alpha_n-\beta_u}{\beta_{\ell}-\beta_u } },
 	\end{align}
 	where $n\in[N]$. Thus, each worker node sends a coded vector of size $\frac{d}{m}$ to the master node.
 	\item
 	{Decoding Phase at the Master:}
 	 Consider the following polynomial function 
 	\begin{align}\label{decoding}
 	{f}(x)=\sum_{\ell=1}^{m}\bigg(\sum_{i=1}^{K}{\mathbf{g}_i[\ell]\prod\limits_{j=1\atop j:\mathcal{D}_i\notin\Gamma_j\hfill}^{N}{\frac{x-\alpha_j}{\beta_{\ell}- \alpha_j}}}\bigg)\prod\limits_{u=1\hfill\atop u\ne\ell\hfill}^{m}{\frac{x-\beta_u}{\beta_{\ell}-\beta_u } }.
 	\end{align}
 	Recall that $\tilde{\mathbf{g}}_n$ is an evaluation of the universal polynomial function ${f}(x)$ at $\alpha_n$, for $n\in[N]$. Since the resulting distributed system tolerates $s$ straggling and $a$ adversarial worker nodes, each data partition at least exists in $r=2a+s+m$ worker nodes. Thus, ${f}(x)$ is a polynomial function of degree $\deg{f}=N-s-2a-1$ with $N-s-2a$ coefficients.
 	According to Reed-Solomon decoding/Lagrange interpolation, to have resistance against $a$ adversarial worker nodes, receiving $\deg{f}+2a+1$ results from worker nodes is necessary \cite{lin2001error}. Thus, by waiting for the results of any $\deg{f}+2a+1=N-s$ non-straggling worker nodes the coefficients of $f(x)$ will be recovered, where at most $a$ results might be erroneous.	
 	\item{Final Result:} The aggregated gradient vector $\mathbf{g}=\sum\limits_{k=1}^{K}\mathbf{g}_k$ can be calculated as follows 
 	\begin{align}
 	\mathbf{g}[\ell]\triangleq\sum_{k=1}^{K}{\mathbf{g}_k[\ell]}={f}(\beta_{\ell}),\hspace{2mm}\text{for } \ell\in[m].
 	\end{align}
 \end{enumerate}

\begin{remark}
	The problem in \cite{ye2018communication} can be covered as a special case of the proposed scheme when the number of data partitions assigned to each worker is equal. That means for each $n\in[N]$ we have $|\Gamma_n|=\gamma$, where $\gamma\in\mathbb{N}$.
\end{remark}
\begin{remark}
	The proposed gradient coding scheme achieves the communication cost of $C^*=\frac{d}{r-2a-s}$.
\end{remark}
\begin{remark}
The computational complexity of the interpolation of a polynomial function of degree $n\hspace{-1mm}-\hspace{-1mm}1$ is $\mathcal{O}(n\log^2 n)$~\cite{li2000arithmetic}. Thus, 	the complexity of decoding in the proposed scheme is $\mathcal{O}((N-s)\log^2(N-s))$.
\end{remark}
\begin{remark}
	The proposed gradient coding scheme can be used to compute any sum of functions using a distributed setting.
\end{remark}

\subsection{Converse proof of Theorem \ref{th1}}\label{converse}
\begin{proof}
	For any linear encoding and general decoding functions we prove $C^*(s,a,\{\mathcal{A}_k\}_{k=1}^K)\ge\frac{d}{r-2a-s}$.  Suppose $(\mathcal{E}_{n\in\mathbb{N}},\mathcal{H})$ can tolerate $a$ adversarial and $s$ straggling worker nodes. Thus, using the results of available worker nodes, the master node can recover $\mathcal{H}\big(\mathcal{F},\{\tilde{\mathbf{g}}_j\}_{j\in\mathcal{F}}\big)=\sum\limits_{k=1}^{K}\mathbf{g}_k$, where $\tilde{\mathbf{g}}_n=\mathcal{E}_n\big( \{\mathbf{g}_i\}_{i: \mathcal{D}_i\in\Gamma_n}\big)$ for $n\in[N]$. Assume if the $n$th worker node is adversarial then it sends $\mathbf{\tilde{g}}_n+\bm{\epsilon}_n$, where $\bm{\epsilon}_n\in\mathbb{R}^d$ is an arbitrary vector. Let us denote the coded vectors, sent by worker nodes, by $\tilde{\mathbf{G}}=(\tilde{\mathbf{g}}_1+\bm{\epsilon}_1,\tilde{\mathbf{g}}_2+\bm{\epsilon}_2,\dots,\tilde{\mathbf{g}}_N+\bm{\epsilon}_N)$, where for a non-adversarial worker node $W_n$ we have $\bm{\epsilon}_n= \mathbf{0}$.
	
	Recall that the $k$th data partition $\mathcal{D}_k$ is assigned to  $|\mathcal{A}_k|$ worker nodes and we define that $r=\min\limits_{k\in[K]}|\mathcal{A}_k|$. Theorem~\ref{th1} is equivalent to $r>2a+s$. The proof goes by contradiction. Suppose for some $j\in[K]$, $\mathcal{D}_j$ is assigned to  $r_j\triangleq|\mathcal{A}_j|\le2a+s$ worker nodes. For example $\mathcal{D}_j$ is replicated in $r_j=2a+s$ worker nodes $W_1,W_2,\dots,W_{2a+s}$.  Now assume $W_1,\dots,W_s$ are straggling worker nodes. Thus, the master node must recover the aggregated gradient vector using the answers from $W_{s+1},\dots,W_{N}$. As a genie-aided argument, assume that 
	$\{\mathbf{g}_k\}_{k=1,\atop{k\ne j} }^{K}$ is already available
	at the master node. Thus, recovering the summation $\sum_{k=1}^K\mathbf{g}_k$ is equivalent to recovering $\mathbf{g}_j$ at the master node. In other words, the value of $\mathbf{g}_j$ can be calculated from $\{\mathbf{g}_k\}_{k=1,\atop{k\ne j} }^{K}$ and the answers of $W_{s+1},\dots,W_{2a+s}$, where there exists $a$ adversarial answers among them. Now assume two different values for $\mathbf{g}_j$ as $\mathbf{g}_j^{(1)}$ and $\mathbf{g}_j^{(2)}$. Thus, for a resilient system consisting of $a$ adversarial and $s$ straggling worker nodes we have
	\begin{align}\label{adv}
	\mathcal{H}_{\text{EGC}}\big(\{W_{s+1},\dots,W_{N}\},
	\tilde{\mathbf{G}}^{(1)}_{s+1:N}\big)&=\mathbf{g}_j^{(1)}+\sum_{k=1\atop k\ne j}^{K}\mathbf{g}_k,\\
	\mathcal{H}_{\text{EGC}}\big(\{W_{s+1},\dots,W_{N}\},
	\tilde{\mathbf{G}}^{(2)}_{s+1:N}\big)&=\mathbf{g}_j^{(2)}+\sum_{k=1\atop k\ne j}^{K}\mathbf{g}_k,
	\end{align}
	where $\tilde{\mathbf{G}}^{(i)}_{s+1:N}=(\tilde{\mathbf{g}}_{s+1}^{(i)}+\bm{\epsilon}_{s+1}^{(i)},\dots,\tilde{\mathbf{g}}_{2a+s}^{(i)}+\bm{\epsilon}_{2a+s}^{(i)},\tilde{\mathbf{g}}_{2a+s+1}+\bm{\epsilon}_{2a+s+1}^{(i)}\dots,\tilde{\mathbf{g}}_N+\bm{\epsilon}_{N}^{(i)})$ for $i=1,2$. Now consider the two scenarios as follows
	\begin{enumerate}
		\item 
		$\bm{\epsilon}_n^{(1)}=\tilde{\mathbf{g}}_{n}^{(2)}-\tilde{\mathbf{g}}_{n}^{(1)}$, for $n=s+1,\dots,s+a$, and $\{\bm{\epsilon}_n^{(1)}\}_{n=s+a+1}^{N}=\mathbf{0}$,
		\item
		$\bm{\epsilon}_n^{(2)}=\tilde{\mathbf{g}}_{n}^{(1)}-\tilde{\mathbf{g}}_{n}^{(2)}$, for $n=s+a+1,\dots,s+2a$, and $\bm{\epsilon}_n^{(2)}=\mathbf{0}$ for $n=s+1,\dots,s+a,s+2a+1,\dots,N$.
	\end{enumerate}
	These two cases are valid scenarios with $a$ adversarial worker nodes,  with the same inputs for the decoding function, thus $\mathcal{H}\big(\{W_{s+1},\dots,W_{N}\},
\tilde{\mathbf{G}}^{(1)}_{s+1:N}\big)=\mathcal{H}\big(\{W_{s+1},\dots,W_{N}\},
\tilde{\mathbf{G}}^{(2)}_{s+1:N}\big)$ or $\mathbf{g}_j^{(1)}=\mathbf{g}_j^{(2)}$. This is a contradiction. Therefore, $\min\{|\mathcal{A}_k|\}_{k=1}^K\ge 2a+s+1$.
	
	Now in terms of communication cost, recall that each worker node sends $C$ symbols to the master node. We aim to prove that $C\ge \frac{d}{r-2a-s}$. Let us assume that the master node receives $d_1$ symbols from $r-2a-s$ non-straggling and non-adversarial worker nodes, i.e., $d_1\triangleq C(r-2a-s)$, where $d_1<d$. For any linear encoding, the coded vector received from those $r-2a-s$  worker nodes can be expressed as follows  
	\begin{align}
		\hat{\mathbf{g}}_{d_1\times1}\triangleq\mathbf{A}_{d_1\times d}\times \mathbf{g}_{k}, \hspace{1cm}\text{for }k\in[K],
	\end{align}
	where $\mathbf{A}$ is the matrix equivalence form of the encoding function. Since $d_1< d$, there exists two different values for $\mathbf{g}_k\in\mathbb{R}^d$ as $\mathbf{g}_k^{(1)}$ and $\mathbf{g}_k^{(2)}$ which are mapped to the one coded vector $\hat{\mathbf{g}}_{d_1\times 1}$. For these two vectors, we have the same issue for other $2a$ non-straggling worker nodes which is argued in \eqref{adv}. As a result,  $\mathbf{g}_k^{(1)}$ and $\mathbf{g}_k^{(2)}$ can not be distinguished by the master node which gives a contradiction. Therefore, $C^*(r-2a-s)\ge d$ or $C^*\ge\frac{d}{r-2a-s}$. This completes the proof of Theorem \ref{th1}.
\end{proof}

 \section{Matrix Polynomial Computation of the Aggregated Gradient Matrix}\label{funcComp}
In this section, the goal is to compute the evaluation of a matrix polynomial $h(\mathbf{X}):\mathbb{V}\to\mathbb{U}$ with aggregated gradient matrix $\sum_{k=1}^{K}\mathbf{G}_k$ as variable, where $\mathbb{V}$ and $\mathbb{U}$ are the set of the real matrices. In other words, the objective is to find the value of $h(\mathbf{G}_1+\dots+\mathbf{G}_K)$, where $\mathbf{G}_k\in\mathbb{R}^{d\times d}$ is the computed partial gradient matrix over data set $\mathcal{D}_k$ for $k\in[K]$. Note that in this context, we focus on polynomial function of square matrix which is a well-defined concept. Thus, a distributed system with one master node and $N$ worker nodes $W_1,\dots, W_N$ is utilized to compute the evaluation of $h(\sum_{k=1}^{K}{\mathbf{G}_k})$ over partitioned data set $\mathcal{D}=\{\mathcal{D}_i\}_{i=1}^K$. We have an arbitrary data placement among worker nodes and data partition $\mathcal{D}_k$ is assigned to the worker nodes in $\mathcal{A}_k$. Also, assume that in the distributed system, there may be at most $s$ straggling and $a$ adversarial worker nodes that send malicious messages to affect the whole computation. The idea of the proposed scheme in Section \ref{scheme}, can be used in a simple way with some changes to calculate the polynomial function of the aggregated gradient matrix.
\begin{theorem}\label{th2}
	Consider the problem of computing $h(\mathbf{G}_1+\dots+\mathbf{G}_K)$ over a cluster of $N$ worker nodes consisting of $s$ straggling and $a$ adversarial nodes, where $h$ is a polynomial function of degree $\deg h$. Assume that partial gradient matrix $\mathbf{G}_k$ is available at worker nodes in $\mathcal{A}_k$, for $k\in[K]$. In this setting, there exists a scheme with communication cost of $C=d^2$ per worker node 
	if  $\min\limits_{k\in[K]}|\mathcal{A}_k|\ge N-\lfloor\frac{N-2a-s-1}{\deg{h}}\rfloor$.
\end{theorem}
In the following we present a scheme for computing $h(\sum_{k=1}^{K}\mathbf{G}_k)$  with $\min\limits_{k\in[K]}|\mathcal{A}_k|= N-\lfloor\frac{N-2a-s-1}{\deg{h}}\rfloor$, and communication cost of $C=d^2$.
\begin{enumerate}
	\item Encoding Phase at the Worker Nodes: The $n$th worker node computes the following coded matrix
	\begin{align}
	\tilde{\mathbf{G}}_n=h\bigg(	\sum_{i=1\atop i: \mathcal{D}_i\in\Gamma_n}^{K}\hspace{-2mm}{\mathbf{G}_i\hspace{-2mm}\prod\limits_{j=1\atop j:\mathcal{D}_i\notin\Gamma_j\hfill}^{N}{\frac{\alpha_n-\alpha_j}{\beta- \alpha_j}}}\bigg),
	\end{align}
	where $\alpha_1,\dots,\alpha_N\in\mathbb{R}$ are $N$ distinct real values, and $\alpha_n$ is assigned to the worker node $W_n$ for $n\in[N]$, and	$\beta$ is a real point, where $\beta\ne\alpha_i$ for $i\in[N]$. 
	\item Decoding Phase at the Master Node: Consider  polynomial function $f(x)$ as
	\begin{align}\label{fx}
		f(x)=\sum_{i=1}^{K}{\mathbf{G}_i\prod\limits_{j=1\atop j:\mathcal{D}_i\notin\Gamma_j\hfill}^{N}{\frac{x-\alpha_j}{\beta- \alpha_j}}}.
	\end{align}
	The master node needs to recover the following polynomial function
	\begin{align}
	h(f(x))= h\bigg(  \sum_{i=1}^{K}{\mathbf{G}_i\prod\limits_{j=1\atop j:\mathcal{D}_i\notin\Gamma_j\hfill}^{N}{\frac{x-\alpha_j}{\beta- \alpha_j}}}\bigg).
	\end{align}
	It is clear that computation result of worker node $W_n$ denoted by $\tilde{\mathbf{G}}_n$, is an evaluation of the polynomial function $h(f(x))$ at point $\alpha_n$, for $n\in[N]$. Data assignment in this scheme is done in a way that having received the results of $N-s$ worker nodes, the master node can recover $h(f(x))$.
	\item Final Result: Matrix polynomial evaluation over the aggregated gradient matrix is calculated as follows
	\begin{align}
	h\bigg(\sum_{k=1}^{K}\mathbf{G}_k\bigg)=h(f(\beta)).
	\end{align}
\end{enumerate}
Now we present the proof of Theorem \ref{th2} by stablishing the correctness of the proposed achievable scheme.
	\begin{proof}
		Suppose each partition of the data set is assigned to at least $r$ worker nodes. According to the structure of $f(x)$ in \eqref{fx}, the degree of $h(f(x))$ is $\deg{h}\times\deg{f}=(N-r)\deg{h}$. Because of the adversarial worker nodes the decoding step needs a slight hesitation. As shown in the proof of Theorem \ref{th1} to tolerate $a$ adversarial worker nodes, the extra replication factor of $2a+1$ is needed. Thus, 
		to recover $h(f(x))$ the computation results of $(N-r)\deg{h}+2a+1$ worker nodes are required, where at most $a$ results are adversarial. Note that each result is a square matrix with size of $d\times d$. To guarantee the robustness against $s$ straggling and $a$ adversarial worker nodes we have
		\begin{align}
		N-s&\ge(N-r)\deg{h}+2a+1,\\
		r&\ge N-\frac{N-2a-s-1}{\deg{h}}.
		\end{align}
		Hence, with   $\min\limits_{k\in[K]}|\mathcal{A}_k|=N-\lfloor\frac{N-2a-s-1}{\deg{h}}\rfloor$ and communication cost of $C=d^2$ per worker node, the proposed scheme can compute the polynomial function of aggregated gradient matrix.
	\end{proof}

%\begin{remark}
%	A problem of computation of any arbitrary function of $\mathbf{G}$ which is not limited to polynomial computations can be considered. The final result can be recovered using approximation methods which are discussed in Section \ref{appSec}.
%\end{remark}
\section{Approximate Gradient Coding}\label{appSec}
Let us consider the achievable scheme of Section \ref{scheme} for the case where $a=0$. Assume that, as the design parameters, (i) the number of stragglers is expected to be less than or equal to $s$ worker nodes,  for some $s \in \mathbb{N}$,  (ii) the communication cost of each worker node must be less than or equal to $R$ symbols, for some $R \in \mathbb{N}$.  Let us assume that we use the scheme of Section \ref{scheme}.  We observe that
\begin{enumerate}
	\item If we want each worker node to send at most $R$ symbols,   we need to have  $R \geq \frac{d}{\min \limits_{k\in[K]} |\mathcal{A}_k| -s}$ or $\min\limits_{k\in[K]} |\mathcal{A}_k| \geq \frac{d}{R}+s$,  where $d$ is the dimension of the gradient vectors.
		\item For decoding scheme to work, we need to have the results of at least $N-s$ worker nodes. 
	\end{enumerate}
	
	In some cases, one or both of these conditions do not hold. In other words,  one or both of the following cases happen: 
	\begin{itemize}
	\item $R$ is too small, i.e. $\min\limits_{k\in[K]} |\mathcal{A}_k| < \frac{d}{R}+s$. 

\item At the time of decoding, more than $s$ worker nodes appear to be stragglers. 
		\end{itemize}
		
		In this case, we cannot use the scheme of Section \ref{scheme}. In this section, we propose an alternative approach that provides us with an approximation of the aggregated gradient vector $\sum_{k=1}^{K}\mathbf{g}_k$, even if both of the above undesirable scenarios happen. 
		
		Approximated gradient coding has been investigated in \cite{grad_app_dimakis,glasgow2020approximate,charles2017approximate,wang2019erasurehead,bitar2020stochastic}. However, the schemes of \cite{grad_app_dimakis,charles2017approximate,wang2019fundamental} focus on the cases where $R=d$ and only condition (1)  may not hold. In addition,  they only consider homogeneous scenarios.  However,  the proposed scheme here can address very general cases. 	
		
		Recall that the decoding method of the proposed scheme in Section \ref{scheme} is based on interpolation of some \emph{polynomial} functions. This is equivalent to a system of linear equations with a Vandermonde matrix \cite{davis1975interpolation}, as the matrix of coefficients. However, Vandermonde matrices over real numbers 
		are ill-conditioned. More specifically, their condition numbers grow exponentially with the size of the matrices \cite{ye2018communication,gautschi1987lower}. This means that it can be computationally unstable. In this section, we leverage the rational functions to overcome some of the deficiencies of polynomial interpolation. Moreover, since the rational functions are a larger class of functions compared with polynomial functions,  it leads to better approximation results for aggregated gradient vector for the cases where one or both of the conditions mentioned before are violated~\cite{berrut2005recent,pachon2012fast}.  
	
		In this approximated computation, each worker node linearly encodes all the partial gradient vectors,  locally available, as $\tilde{\mathbf{g}}_n=\mathcal{E}_n\big( \{\mathbf{g}_i\}_{i: \mathcal{D}_i\in\Gamma_n}\big)$ and sends to the master node, where $\mathcal{E}_n$ is the encoding function that maps the raw partial gradient vectors to the coded partial gradient $\tilde{\mathbf{g}}_n$ for the $n$th worker node. The results of an arbitrary subset of fastest worker nodes $\mathcal{F}$ is then used by the master node to  approximately recover the aggregated gradient $\mathbf{g}$, i.e.,  $\mathbf{g}\approx\mathcal{H}_{\text{AGC}}\big(\{\tilde{\mathbf{g}}_j\}_{j\in\mathcal{F}},\mathcal{F}\big)$, where $\mathcal{H}_{\text{AGC}}$ is a decoding function used in the approximate gradient coding scheme.
\subsection{Preliminaries}
Before introducing the approximated scheme,  we need some preliminaries.
\begin{definition}[Interpolation \cite{Cheney}]
	Let the set of $n$ distinct interpolation points as $\mathcal{X}_n=\{x_i\}_{i=0}^n$ be given in a closed interval $[a,b]$ with evaluations of a real value function $f\in \mathcal{C}[a,b]$ at these points, i.e., $f_i=f(x_i)$ for $i\in[0:n]$, where $\mathcal{C}[a,b]$ is the space of all continuous functions on $[a,b]\subset\mathbb{R}$. An interpolation method can be defined as follows
	\begin{align}\label{inter}
	\tilde{f}(x)=\sum_{i=0}^{n}f_i\varphi_i(x),
	\end{align}
	where $\varphi_i(x)$ for $i\in[0:n]$ are the basis functions of the interpolant method which do not depend on $f_i$. 
\end{definition}
\begin{remark}
	Note that different basis functions lead to different interpolation methods. Thus, choosing the basis functions and interpolation points affect the error of the interpolation method.  \textit{Lebesgue constant}  is
	one of the best criteria to determine which sets of interpolation points are appropriate.
	
\end{remark}
\begin{definition}[Lebesgue Constant \cite{berrut1997,Cheney}]
	Consider the basis functions $\mathcal{B}_n=\{\varphi_i\}_{i=0}^n$ and $n$ distinct interpolation points $\mathcal{X}=\{x_i\}_{i=0}^n\in[a,b]$.  The goal is to find a linear interpolant $\tilde{f}$ as in \eqref{inter} from an $n$-dimensional vector space $\mathbb{A}$. Lebesgue constant can be expressed as follows 
	\begin{align}\label{Lambda}
	\Lambda_n= \max_{x\in[a,b]}\sum_{i=0}^{n}{|\varphi_i(x)|},
	\end{align}
	which is derived from the theory of linear interpolation operators.
\end{definition}
One of the applications of the Lebesgue constant is bounding of the approximation error in terms of the best approximation in the corresponding $n$-dimensional vector space $\mathbb{A}$. In other words, we have \cite{shortCourse}
\begin{align}\label{error}
\norm{f-\tilde{f}}\le (1+\Lambda_n)\norm{f-r^*},
\end{align}
where $r^*$ is the best approximation in $\mathbb{A}$ of function $f$. According to \eqref{error}, having a better interpolation of $f$ in a particular vector space depends on choosing properly distributed interpolation points that cause the smaller Lebesgue constant.  

Since rational polynomials use a larger class of functions, the resulting approximations are much better than polynomial interpolations \cite{berrut2005recent,pachon2012fast}. In the following a specific representation for the rational interpolation is defined.
\begin{definition}[Berrut's Rational Interpolation \cite{berrut1988rational}]
	The following rational function 
	\begin{align}\label{berrut}
	\tilde{f}_{\text{Berrut}}(x)=\sum_{i=0}^{n}\frac{{\frac{{(-1)}^i}{(x-x_i)}}}{\sum_{j=0}^{n}\frac{{(-1)}^j}{(x-x_j)}}f_i,
	\end{align}
	is called \textit{Berrut's Rational Interpolant} which interpolates $f_k$ at $x_k, k\in[0:n]$. The basis functions of this interpolant is denoted by
	\begin{align}
	\varphi_{i,\text{Berrut}}=\frac{\frac{{(-1)}^i}{(x-x_i)}}{\sum_{j=1}^{n}\frac{{(-1)}^j}{(x-x_j)}}, \hspace{1cm}\text{for }i\in[0:n].
	\end{align}
\end{definition}
In \cite{carnicer2010weighted,wang2010rational}, it is shown that fixed interpolation points could have different effects on two different interpolation methods. For example, 
using the equidistant points in the corresponding closed interval for rational interpolations causes a much smaller Lebesgue constant than in polynomials. In the following, we introduce a general class of interpolation points having reasonable effects on Berrut's rational interpolation. 
\begin{definition}[Well-Spaced Points \cite{bos2013bounding}]\label{wellSpaced}
	Let $\mathcal{X}_n=\{x_i\}_{i=0}^{n}$ be a set of ordered distinct interpolation points. Consider a family of sets, i.e., $\mathcal{X}=(\mathcal{X}_n)_{n\in\mathbb{N}}$, if there exist constants $\rho,\nu\ge1$ such that the following conditions 
	\begin{align}
	(1)\hspace{0.5cm} &\frac{x_{k+1}-x_k}{x_{k+1}-x_j}\le\frac{\rho}{k+1-j},&&\text{ for }j\in[0:k],k\in[0:n-1],\label{cod1}\\
	(2)\hspace{0.5cm}&\frac{x_{k+1}-x_k}{x_j-x_k}\le\frac{\rho}{j-k}&&\text{ for } j\in[k+1:n],k\in[0:n-1],\label{cond2}\\
	(3)\hspace{0.5cm}&\frac{1}{\nu}\le\frac{x_{k+1}-x_k}{x_k-x_{k-1}}\le \nu,&&\text{ for }k\in[n-1]\label{cond3},
	\end{align}
	are held, then $\mathcal{X}=(\mathcal{X}_n)_{n\in\mathbb{N}}$ is called a family of well-spaced points. Note that $\nu$ and $\rho$ must be independent of $n$.
\end{definition}

In \cite{bos2013bounding}, it is shown that if $\mathcal{X}_n=\{x_i\}_{i=0}^n$ is used as a set of ordered interpolation points and it is from a family of well-spaced points, then  the Lebesgue constant in Berreut's rational interpolation is bounded as follows 
\begin{align}
\Lambda_n\le(\nu+1)(1+2\rho\ln n).
\end{align}

\subsection{The Proposed Scheme}
In this subsection, we introduce the proposed approximate gradient.
This approximated computation has three main steps as follows.
\begin{enumerate}
	\item Encoding Phase at the Worker Nodes: Let us assume that the communication cost per worker node is limited to $R$, for some $R\in\mathbb{N}$. For encoding, follow the encoding of Section \ref{scheme} with $m=\lfloor \frac{d}{R}\rfloor$.
	\item
	Decoding Phase at the Master Node:  Having received coded partial gradient vectors from a subset of
	worker nodes denoted by $\mathcal{F}$, the master node can approximately compute the aggregated gradient vector $\mathbf{g}$ in a numerically stable manner. In this scheme, the master node uses Berrut's rational interpolant to approximately interpolate the universal polynomial function ${f}(x)$ as follows
	\begin{align}\label{app}
	\tilde{{f}}_{\text{Berrut},\mathcal{F}}(x)=\sum_{i=1}^{|\mathcal{F}|}\frac{{\frac{{(-1)}^i}{(x-\tilde{\alpha}_i)}}}{\sum_{j}\frac{{(-1)}^j}{(x-\tilde{\alpha}_j)}}{f}(\tilde{\alpha}_i),
	\end{align}
	where $\tilde{f}(x)$ interpolates $f(x)$ at $\tilde{\alpha}_i$ for $i\in[|\mathcal{F}|]$, where $\{\tilde{\alpha}_i\}_{i=1}^{|\mathcal{F}|}=\{\alpha_i,i\in\mathcal{F}\}\subset\{ \alpha_j\}_{j=0}^N$ are the interpolation points. 
	\item Final Result: The master node then approximately computes  $\mathbf{g}$ as follows
	\begin{align}
	\mathbf{g}[\ell]=\sum_{k=1}^{K}{\mathbf{g}_k[\ell]}\approx\tilde{{f}}_{\text{Berrut},\mathcal{F}}(\beta_{\ell}),\hspace{2mm}\text{for } \ell\in[m].
	\end{align}
\end{enumerate}
\begin{remark}
	As mentioned before, there is no strict notion of the minimum number of required outcomes from worker nodes. In other words, $|\mathcal{F}|$ can be less than $N-s$ and by receiving more coded partial gradient vectors form worker nodes, the aggregated gradient vector can be calculated more accurately. 
\end{remark}
\begin{remark}
	It is assumed that in the proposed approximate gradient coding scheme, there is no adversarial worker node among other nodes.
\end{remark}
\begin{remark}
	Let $\mathcal{X}_{N-s_1}=\{\alpha_i\}_{i=1}^{N-s_1}$ be a set of ordered distinct values of non-straggling worker nodes which are considered as interpolation points for decoding step, where $s_1$ is grater than the expected value for the number of stragglers, i.e., $s_1>s$. Assume $\mathcal{X}_{N-s_1}$ is a subset of Chebyshev points of second kind, i.e.,  $\mathcal{X}_{N-s_1}\subset\{\cos{\frac{k\pi}{N}}\}_{k=0}^N$.
	In \cite{jahani2020berrut}, we proved that there are some constant parameters $\rho,\nu\ge1$ such that the three conditions \eqref{cod1}, \eqref{cond2} and \eqref{cond3} hold for each $x_i\in \mathcal{X}_{N-s_1},i\in[N-s_1]$. Therefore, $\mathcal{X}_{N-s_1}$ contains well-spaced points. In \cite{jahani2020berrut}, we also showed that the Lebesgue constant for Berrut's rational interpolant grows logarithmically in the size of  \emph{a subset of Chebyshev points}. Thus, in this scheme, we suggest to choose  $\alpha_j, j\in[N]$,  as Chebyshev points of the second kind.
\end{remark}
\begin{remark}
	Since $f(x)$ has a continuous
	second derivative on $[-1,1]$, the approximation error of
	$\tilde{{f}}_{\text{Berrut},\mathcal{F}}(x)$  defined in \eqref{app}  is upper bounded as follows
	\begin{align}\label{bounded}
	\norm{\tilde{{f}}_{\text{Berrut},\mathcal{F}}(x)-f(x)}\le2(1+\nu)\sin\big({\frac{(s_1+1)\pi}{2N}}\big)\norm{f^{\prime\prime}(x)},
	\end{align} 
	if $N-s_1$ is odd, and if $N-s_1$ is even we have
	\begin{align}\label{bounded2}
	\norm{\tilde{{f}}_{\text{Berrut},\mathcal{F}}(x)-f(x)}\le2(1+\nu)\sin\big({\frac{(s_1+1)\pi}{2N}}\big)\bigg(\norm{f^{\prime\prime}(x)}+\norm{f^{\prime}(x)}\bigg),
	\end{align}
	where $s_1$ and $N$ are the number of stragglers in the running time and the total number of worker nodes, respectively. Also, $\nu=\frac{(s_1+1)(s_1+3)\pi^2}{4}$, where $s_1< N-2$ \cite{jahani2020berrut}. Thus, the error of the proposed approximate gradient coding scheme can be bounded using the inequalities in \eqref{bounded} and \eqref{bounded2}. 
\end{remark}
\begin{remark}
	According to \eqref{app}, 
	the decoding function which is based on Berrut's rational interpolation has the computational complexity of $\mathcal{O}(|\mathcal{F}|)$, where $\mathcal{F}$ is the set of fastest non-straggling worker nodes from which the master received outcomes.
\end{remark}
\begin{remark}
 In\cite{mascarenhas2014backward}, it is shown that Berrut's rational interpolation is backward stable when the Lebesgue constant $\Lambda_n$ remains small. We call
 interpolation $\tilde{f}$ of function $f$ is $\delta$-\textit{backward stable} if 
  for some small  \textit{backward error} $\delta>0$, and for any $x\in\mathbb{R}$, there exist some $|\delta_x|\le\delta$ such that 
 $\tilde{f}(x)=f(x+\delta_x)$.
\end{remark}
\section{Conclusion}
 In this paper, we considered the problem of gradient coding in heterogeneous distributed systems consisting of straggling and adversarial worker nodes.
For any arbitrary data placement in worker nodes, we characterized the optimum trade-off among communication cost, the number of straggling and adversarial worker nodes, and the minimum number of repetition in computations. The proposed scheme is also extended to the computation of a polynomial function of the aggregated gradient matrix. Finally, we proposed an approximate gradient coding scheme for the cases when the repetition in computations  does not support the restriction on the communication cost or when the number of stragglers  appears to be more than what we designed the system for.

\bibliographystyle{ieeetr}
\bibliography{References}

\end{document}